**Measuring ultra-short pulse widths before and after the objective with a home built autocorrelator**

**Authors:** Sanjeev Kumar Kaushalya and Hans Fried (Light Microscope Facility (LMF) – Core Research Facilities and Services (CRFS), German Centre for Neurodegenerative Diseases (DZNE), Bonn, Germany)

**Abstract:**

In multiphoton microscopy, measurements and compression of the laser pulse width at the setup plays an important role in optimizing the fluorescence output with minimal laser power applications. For example, in deep two and three-photon intravital microscopy it is very crucial to have shortest laser pulses for efficient fluorescence excitation with lowest possible laser power at the sample. There are several published experimental methods and techniques for measuring the group delay dispersion and pulse width. Yet, there are practical limitations for their implementation and beam optimization at intravital setups in individual labs or microscope facility. Here, we present a method to construct a portable optical collinear autocorrelator, to probe the beam directly after the objective and at desired place in the beam path to measure the pulse width, and to optimize the pulse width with a pulse compressor. Finally, we demonstrate that the method increases fluorescence excitation efficiency in three-photon microscopy.

**Introduction:**

Multi-photon microscopy is one of the main imaging methods for deep tissue in-vivo imaging. the recent advancements in the techniques and availability of lasers with high per pulse energy made it possible to routinely perform even deeper imaging by three-photon excitation with far infrared-pulsed lasers (1200-2000 nm). The average rate of fluorescence emission per fluorophore molecule is proportional to [1-3]

$$Fluorescence \propto \frac{\sigma_n}{R^{(n-1)}\tau^{(n-1)}}\left(\frac{\lambda_n P_{av}}{\pi\omega_0^2 hc}\right)^n \tag{1}$$

Where $n$ is the order of excitation, $\sigma_n$ is the n-photon absorption cross section, $\lambda_n$ is the center excitation wavelength, $P_{av}$ is the average laser power, $R$ is the repetition rate of the laser pulses, $\tau$ is the pulse width, $\omega_0$ is the beam waist at the focus, $h$ is Planck constant, and $c$ is speed of light. From Eq (1) the three-photon excitation is quadratically inversely proportional to the pulse width and repetition rate of the laser ($Fluorescence \propto \frac{1}{R^2\tau^2}$). As we know, short pulses of light cannot be monochromatic [4] and shorter laser pulses will always have a larger spectral bandwidth. Therefore dispersion will affect them more and broaden the pulse width significantly. The chromatic dispersion for a laser pulse is quantitatively measured in terms of Group delay dispersion (GDD). In any imaging setup the excitation laser goes through many dispersive media before it gets focused at the sample by the objective. The objectives, especially with large NA, are the major sources of dispersion along with the scan and tube lenses and the adaptive power control optics [2, 5]. A broadened pulse at the sample results in much lower fluorescence and usage of higher laser power to increase the fluorescence signal results quickly in photo damage[6]. For example, if a 50 fs pulse becomes 100 fs the fluorescence is reduced to one fourth.

And to get the same fluorescence it will require 58% more average power. This becomes more and more important in three-photon microscopy as the cross section for three-photon excitation is much smaller than the cross section for two-photon excitation[7, 8] . Thus, three-photon excitation needs higher energy per pulse which in turn strongly increases the probability for photo damage [9, 10]. Precisely controlling the pulse width is therefore important for minimal excitation power and to avoid photo damage.

Ultrafast lasers used in multiphoton microscopy, have very short pulses (50-150 fs) and it is not possible to record the pulse with or pulse profile with fast photodetectors and plot it on an oscilloscope. There are several methods to measure the pulse profile and each have their own advantage and disadvantage. Among those are various time frequency methods, streak camera measurements, spectral interferometry, and autocorrelation methods [11]. Appropriate methods are used depending on the information one needs and other practical considerations like portability, cost, and ease of using. Some of these methods have been used to measure short laser pulses in microscope setups. In particular, group delay dispersion (GDD) and pulse width of the excitation laser at the objective focus have been measured and used to optimize the fluorescence emission [2, 5, 12]. Unfortunately, these methods are mostly applied by experts who are well verse in optics handling and electronics. Many scientists and imaging core facilities actually applying deep tissue in-vivo imaging in their experiments lack this expertise in optics handling and therefore may not be able to optimize their setups for the least power usage.

In this manuscript, we present a design for an autocorrelator, adaptable for 650-1700 nm wavelength range to monitor the pulse width at the sample plane after the objective. Our autocorrelator does not require any modification of the beam path. By using a suitable mirrors placed either before the microscope or after the objective, laser pulse widths can be measured anywhere along the beam path. In addition, we provide a workflow, to construct, analyze, and interpret the data for optimizing the pulses by dispersion compensation. This autocorrelator can easily be installed at any existing home built or commercially available multi-photon microscope, equipped with controllable micrometer stage and z-drive. This should be easy enough to be managed by regular multiphoton users with some experience of beam alignment.

**Material and methods:**

**Construction:**

We constructed the autocorrelator based on detection of fluorescence generated by a two-photon absorption process (second order autocorrelation see supplementary information). The autocorrelator is built completely on a portable 30x40 cm breadboard, which easily accommodate all the components. It can be fixed on an optical bench or on a xy stage of an upright in-vivo imaging microscope, which provides at least 20 cm space below the objective lens. We constructed the autocorrelator with components from Tholabs, PI, and an oscilloscope (Tecktronics) to visualize and record the autocorrelation signals. A detailed step by step construction guide, a complete parts list, and alignment procedure can be found in the supplementary material.

**Pulse width measurements:**

The interferometric autocorrelation-based method is widely used and sufficient for measurements at multiphoton microscopy setups and their optimization [2]. This method is based on the Michelson interferometer. In brief, the laser pulse under consideration is split into two equal and directionally orthogonal pulses (with a 50/50 beam splitter), a reference beam and a beam where the beam path length is changed periodically. Both beams are combined again collinearly to generate interference signal (Figure 1A). Due to the periodically changing path lengths of one beam the interference signal will also be changing periodically. Plot of interference signal versus change in path length generates the interferogram and finally, the interferogram can be used to calculate the pulse width.

One of main advantage of our design is the ease of measurements after the objective. It does not need the insertion of the autocorrelator in the beam path before the microscope and alignment of the interfering beams through the microscope optics or any changes on the optical table. By replacing the first coupling mirror (M1) in Figure 1B by a short focal length off axis parabolic mirror and matching the focus spot of the objective with the focus spot of the parabolic mirror gives a collimated reflected output beam after the parabolic mirror (Figure 1C). The focus spots of the objective and parabolic mirrors are finely adjusted with the help of XY micrometer stage and Z drive available at the microscope. This collimated beam is aligned to the autocorrelator by steering the beam with parabolic mirror and the mirror M2 (Figure 1).

The calibrations and measurements with the autocorrelator are described in detail in the supplementary information. In brief, the beam path in the autocorrelator has to be finely adjusted to obtain an interferogram, which is symmetrical and is having a maximal signal to background ratio close to 8:1. The autocorrelator has two readouts. One is the photo detector current (signal 1) based on the two-photon absorption and the other one is a position readout of the movable mirror which generates the periodic changes of the beam path length (signal 2). For visualizing the interferogram we plot the signal 1 versus signal 2. Signal 2 has to be converted first from delay in length to delay in time. For pulse width measurements we used a envelop fitting routine to plot the upper envelope of the interferogram and determined the full width at half maximum (FWHM). To calculate the pulse width we devided the FWHM by 1.66 and 1.9 for gaussian or sech pulses, respectively. For a highly chirped pulse with dispersive media the profile of the autocorrelation curve will be significantly different and actual pulse width will be much larger even though the spectral bandwidth and the coherence length will be the same. For better estimate of highly chirped pulses we used the intensity autocorrelation deduced from the interferograms. The FWHM values were divided by 1.414 or 1.543 for gaussian and sech pulse, respectively.

**Dispersion compensation**

For pulse width optimization and dispersion compensation we used a prism pair based pulse compressor (APE, FemtoControl specifically meant for 1300-1900 nm wavelength range) before the microscope (Figure 2). The microscope used here was a three-photon microscope built with a Thorlabs Bergamo multiphoton setup and a SPIRIT and NOPA laser solution from SpectraPhysics (Newport, SPIRIT 1030-70 and NOPA-VISIR+I). Dispersion compensation was adjusted under visual control of the interferogram measured after the objective. For 1300 nm negative chirp and for 1700 nm, positive chirp was systematically adjusted as per the manufacturer instructions.

**Results and discussion:**

To analyze GDD in our three-photon setup, we measured the pulse width of the beam before it enters the microscope scanning mirrors (position-I), just before the objective (position-II), and after the objective (position-III) at two wavelengths (1300 nm and 1700 nm). The data from the measurements reveal the dispersion and their respective contribution for GDD at the different elements along the beam path between position I and II (coated mirrors, scan lens, tube lens, and a dichroic mirror) and between position II and III (the objective) (Figure 3).

Figure 3A shows representative interferograms and deduced intensity correlations for the measurements at 1300nm and 1700nm for position-I. Measured pulse width at 1300 nm and 1700 nm are 49.91 fs and 50.61 fs, respectively. At position-I pulses are near transform limited as we can see from the shape of interferogram. In addition, the pulse width measurements from the interferogram and from the intensity correlation are closely matching each other. For the interferogram the peak signal to background ratio is approximately 8:1 and fringes are well resolved. For the intensity autocorrelation peak signal to background ratio is approximately 3:1. Together the data demonstrates that our autocorrelator is sturdy and the setup is free of mechanical vibrations.

Next, we measure the pulse width at position II. At this position the beam has passed through the scan mirrors, scan lens, and tube lens and a dichroic mirror. These optical elements are specifically coated for 1300-1900 nm wavelength range. Measurements at this point show the sum-effect of any dispersion introduced by all these elements at respective wavelengths. Pulse width at 1300 nm and 1700 nm are 114.67 fs and 87.42 fs, respectively (Figure 3B). From the plot we can see the effect of dispersion on the interferogram shape and changes in the pulse width. The pulses are no more transform limited, showing that the microscope optics is affecting pulses at both wavelengths. The GDD contribution from the microscope optics is 1850 fs$^2$ at 1300 nm and -1300 fs$^2$ at 1700 nm.

Finally, we measure the pulse width at position III with the high NA Olympus 25X objective (XLPLN25XWMP2) in place. The objective specifications claim a >70% transmission in the 1300-1700 nm range. Pulse width at 1300 nm and 1700 nm are 200.61 fs and 102.22 fs, respectively (Figure 3C). From the interferogram it is very clear that the objective is a big source of GDD at the tested wavelengths. The pulse width after the objective is two times (1700 nm) or four times (1300 nm) larger compared to the pulse width before the microscope. The total GDD contribution from the microscope and the 25X objective is 3500 fs$^2$ at 1300 nm and -1620 fs$^2$ at 1700 nm.

For dispersion compensation and pulse compression, we started with the prism configurations which compensates for the measured GDD. And while looking at the interferogram at the sample plane we fine adjusted the prism position to get the lowest pulse width possible. With appropriate compensation the interferogram after the objective resembles the interferogram after the laser (Fig 3). Fine pulse optimization is carried out and verified with a fluorescent sample (for example Convalaria rhizome). Prisms are adjusted to achieve the maximum fluorescence intensity and this is also practical procedure for daily usage to check for pulse compression. Figure 4 shows the interferogram and intensity correlation plots after the objective with (Figure 4B) and without GDD compensation (Figure 4A) at 1300nm. After compensation we achieve 56.66 fs pulse width. Images in Figure 4 were taken with the same laser power showing a dramatic fluorescence intensity increase of about 10fold. Form Eq(1) it implies that with the same power, theoretically we will get 12.5 times more fluorescence, thus theoretical and measured values closely resemble each other. Alternatively, at the same fluorescence intensity we can use 2.3 times less power. This will considerably limit the thermal heating and photo damage of the sample during imaging.

**Acknowledgements:**

We thank Eugenio Fava (CRFS, DZNE Bonn) for his great support and for providing the possibility to work on the topic. We also thank Martin Fuhrmann (DZNE Bonn) for helpful discussions on 3P microscopy. This work was supported by the DZNE.

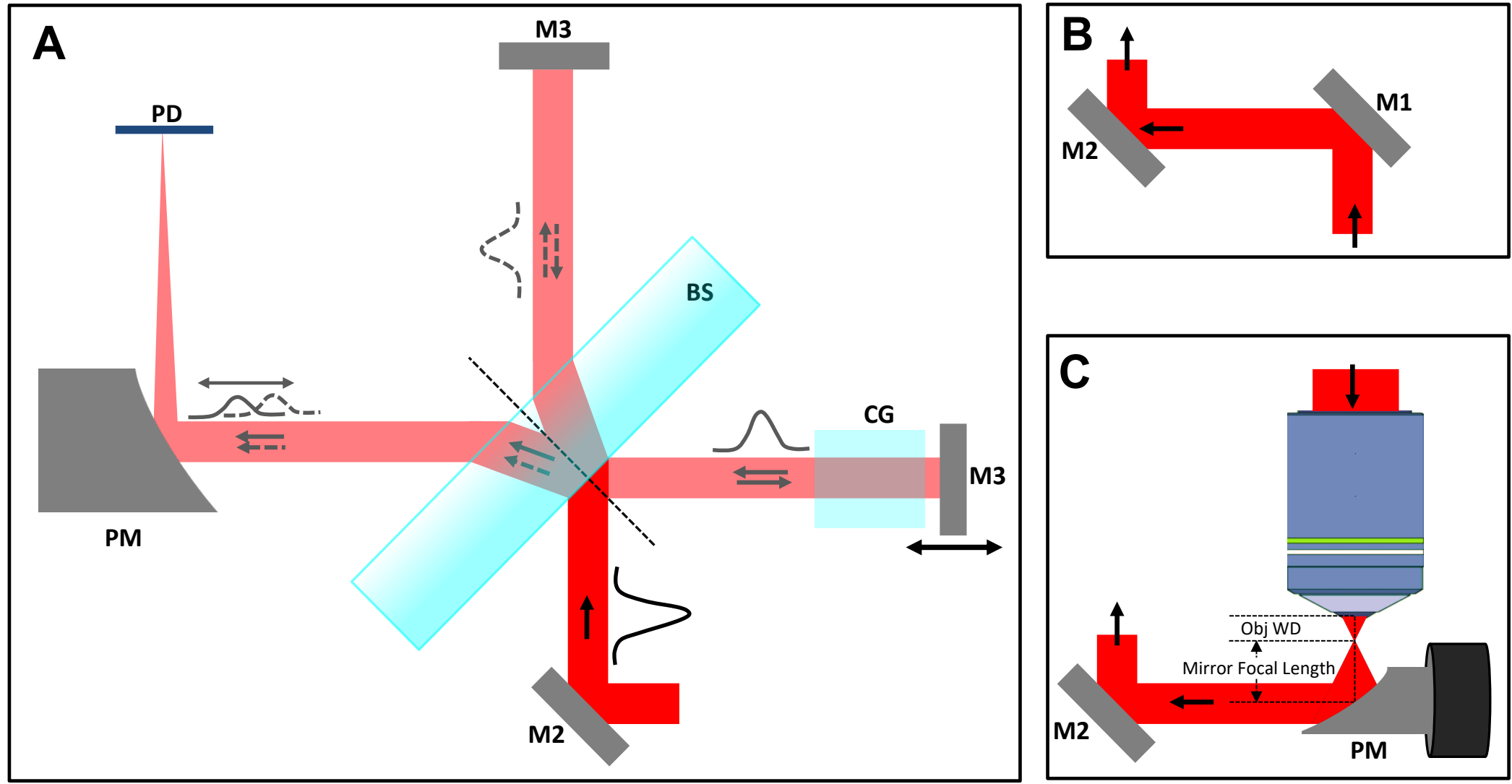

**Figure 1:** A, Schematic diagram of the autocorrelator and its components. The beam splitter is shown much thicker to illustrate the beam path. B, Beam coupling for pulse width measurement after the objective. 90° off axis parabolic mirror helps to achieve a collimated beam for coupling. C, Beam coupling for pulse width measurement before microscope. BS: Beam splitter, CG: Compensation glass, M: Mirror, PD: Photo detector, PM: 90° off axis parabolic mirror

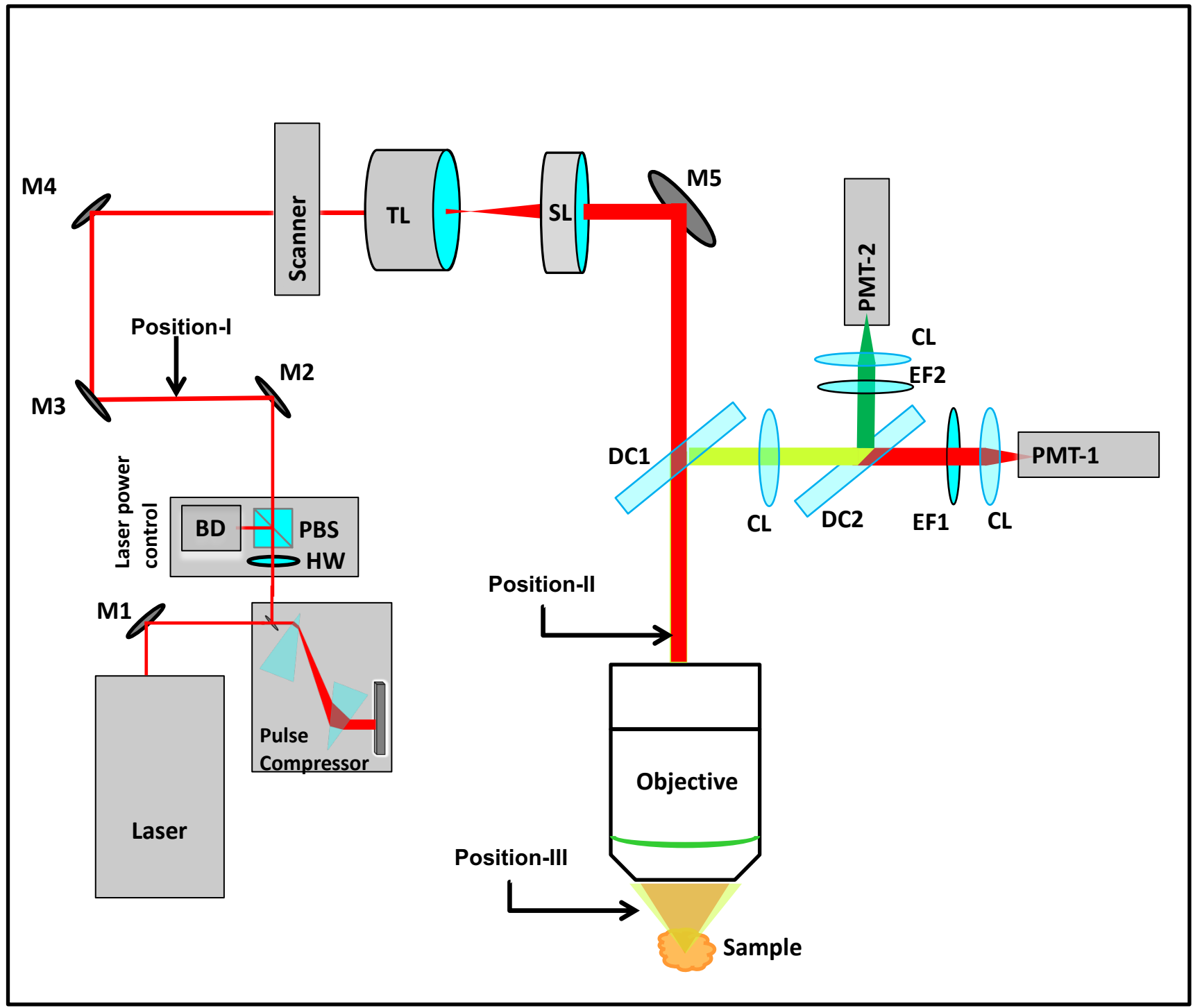

Figure 2: Lay out of the imaging setup. BD: Beam dump CL: Collection Lens, DC: Dichroic, EF: Emission Filter, HW: Half wave plate M: Mirror, PBS: Polarizing beam splitter, SL: Scan Lens, TL: Tube Lens

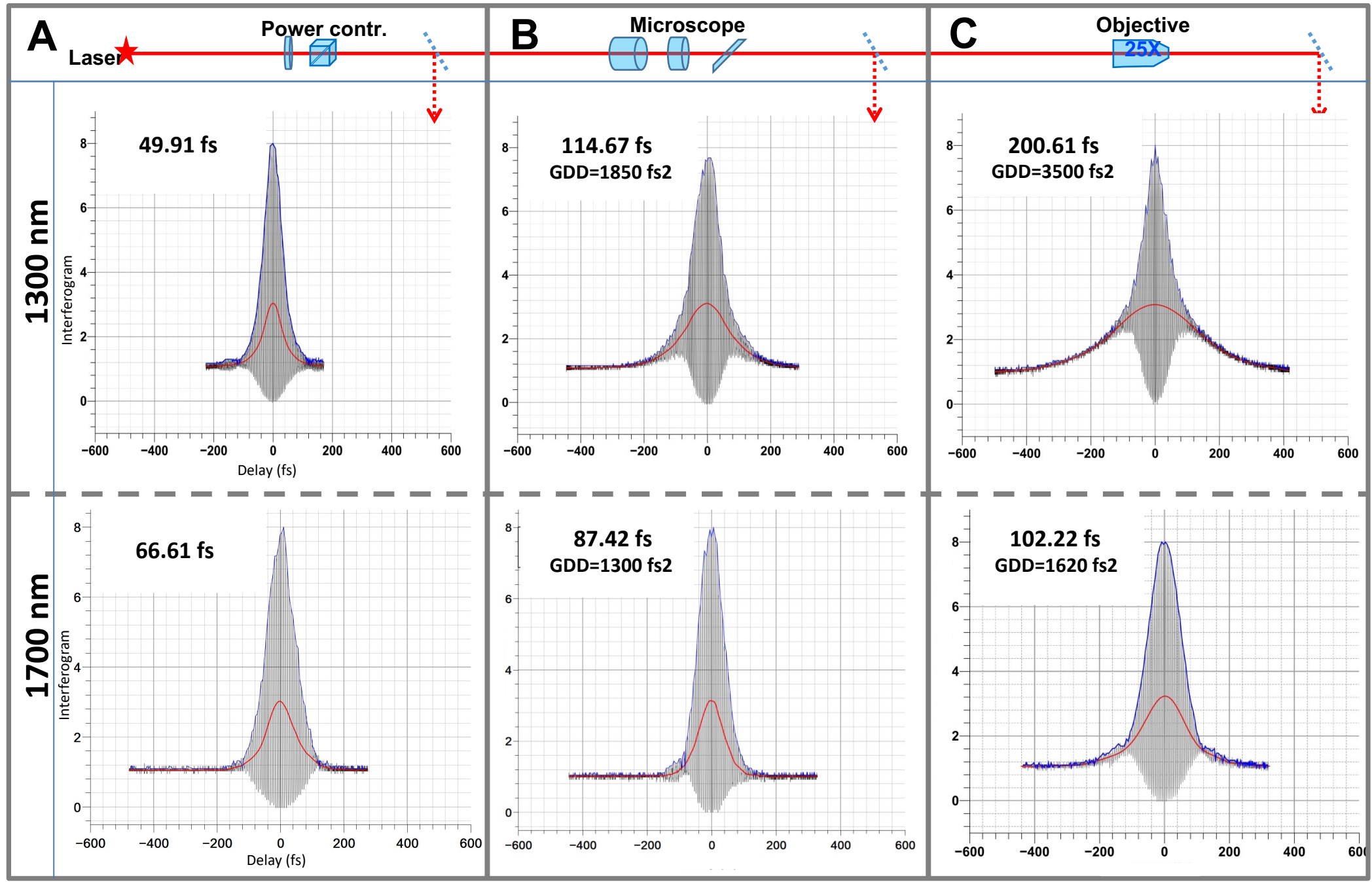

Figure 3: Pulse width measurements at different locations of beam path at the microscope. A, before microscope. B, after microscope optics but before objective. C, after objective. Grey lines, interferogram; red lines, intensity correlation; blue lines, envelope interferogram.

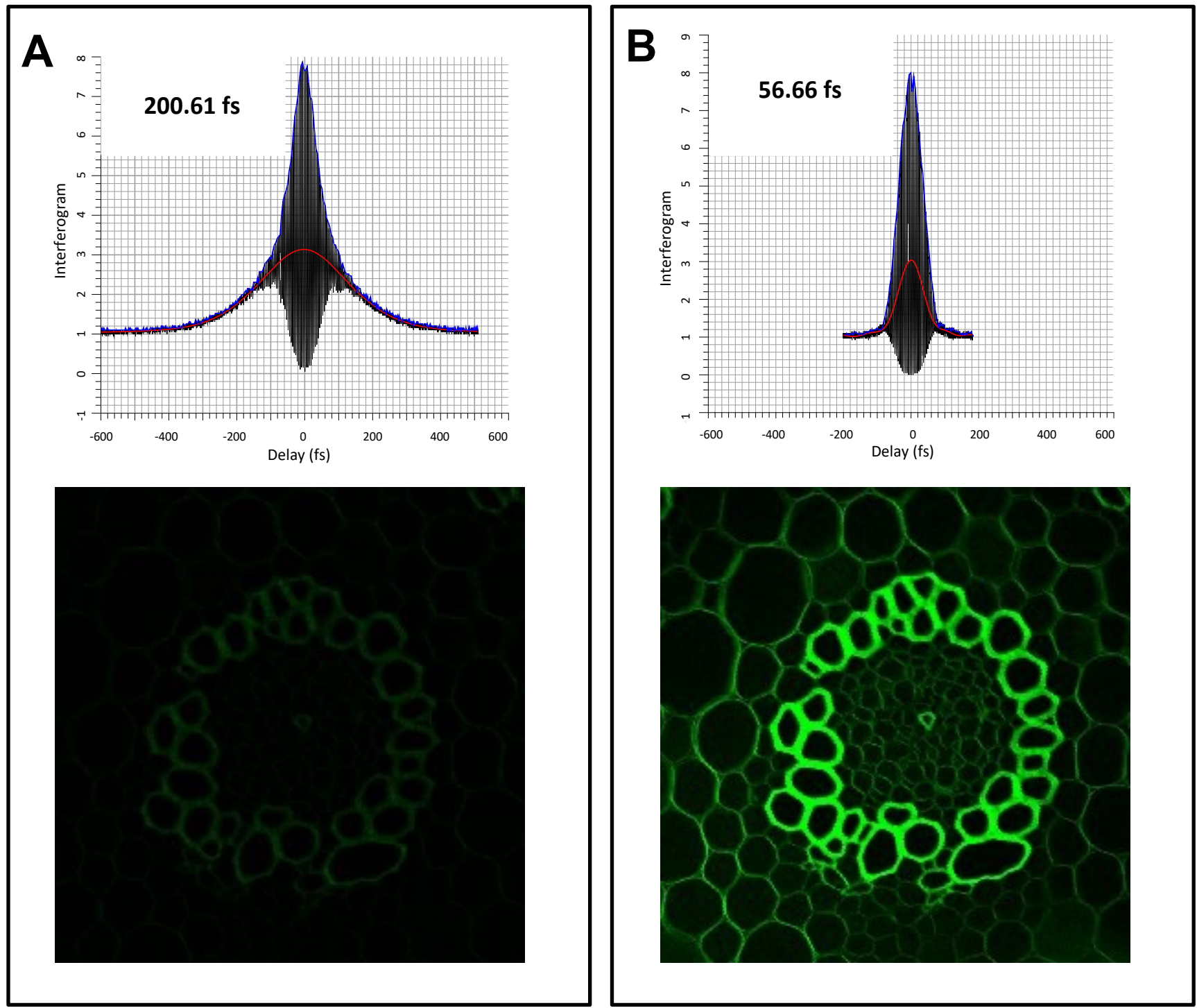

Figure 4: effect of dispersion compensation on Images and pulse width. A, Pulse width and image before compensation. B, Pulse width and image after compensation an increase of 10.2 fold in the fluorescence with same average power. Grey lines, interferogram; red lines, intensity correlation; blue lines, envelope interferogram.